\documentclass[prl,twocolumn]{revtex4}
\usepackage{amsfonts}
\usepackage{mathrsfs}
\usepackage{amsmath}
\usepackage{amssymb}
\usepackage[medium]{titlesec}
\usepackage{bm}
\usepackage[normalem]{ulem}
\usepackage{extarrows}
\usepackage{slashed}
\usepackage{isodateo}
\usepackage{graphicx} 
\usepackage{xcolor}
\usepackage[bookmarksnumbered=true,bookmarksopen=true]{hyperref}
 \hypersetup{colorlinks,%
             linkcolor=[rgb]{0,0.3,0.6}, %
             citecolor=[rgb]{0,0.3,0.6}, %
             urlcolor=[rgb]{0,0.3,0.6}}%
\renewcommand{\FR}[2]{\displaystyle\frac{\,{#1}\,}{#2}}
\newcommand{\fr}[2]{\mbox{$\frac{\,{#1}\,}{#2}$}}
\newcommand{\n}{\nonumber}

\graphicspath{{fig/}}

\def\bge{\begin{equation}}
\def\ede{\end{equation}}
\def\bga{\begin{aligned}}
\def\eda{\end{aligned}}
\def\bgp{\begin{pmatrix}}
\def\edp{\end{pmatrix}}
\def\bgs{\begin{subequations}}
\def\eds{\end{subequations}}
\newcommand{\order}[1]{\mathcal{O}({#1})}
\def\di{{\mathrm{d}}}

\def\pd{\partial}

\def\la{\langle}\def\ra{\rangle}

\def\to{\rightarrow}

\def\ga{\gamma}

\def\ep{\epsilon}

\begin{document} 

\title{Eccentricity Without Measuring Eccentricity: Discriminating Among Stellar Mass Black Hole Binary Formation Channels}

\author{Lisa Randall}
\affiliation{Department of Physics, Harvard University, 17 Oxford St., Cambridge, MA 02138, USA} 

\author{Zhong-Zhi Xianyu}
\affiliation{Department of Physics, Harvard University, 17 Oxford St., Cambridge, MA 02138, USA}  
 
\begin{abstract}

We show how the observable number of binaries in LISA is affected by eccentricity through its influence on the peak gravitational wave frequency, enhanced binary number density required to produce the LIGO observed rate, and the reduced signal-to-noise ratio for an eccentric event. We also demonstrate how these effects should make it possible to learn about the eccentricity distribution and formation channels by counting the number of binaries as a function of frequency, even with no explicit detection of eccentricity. We also provide a simplified calculation for signal-to-noise ratio of eccentric binaries. 
\end{abstract}

\maketitle

\emph{Introduction.} The LIGO/Virgo detections of coalescing black hole binaries (BBHs) marked the dawn of gravitational-wave (GW) astronomy \cite{LIGOScientific:2018mvr}. With increasing statistics from the ongoing run of LIGO/Virgo, we expect to learn many properties of stellar-mass BBHs. One of the important open questions is the formation channel of these merging black hole pairs.

Observing BBHs at a lower frequency in the millihertz range with LISA could provide powerful information about the formation channel. One aspect of LISA measurements that has been considered is the orbital eccentricity of BBHs \cite{Nishizawa:2016jji,Nishizawa:2016eza}. Isolated BBHs typically process little eccentricity while dynamically formed BBHs could have observably large eccentricity. Therefore measuring eccentricity at LIGO could in principle be a good way to differentiate among formation channels.

However, GWs tend to circularize the binary's orbit so that the eccentricity gets reduced together with the orbital size. By the time the BBH enters the LIGO band, the eccentricity is typically small even for dynamically formed binary systems. If possible, it would therefore be better to measure eccentricity at a lower frequency. 

In this paper, we show that LISA should distinguish dynamically formed channels from isolated mergers. Unlike LIGO binaries, BBHs in LISA are in their early inspiral stage and most of them do not chirp significantly, so the emitted GWs from these BBHs stay within a small range of frequencies over the entire LISA mission. The dynamical channels are likely to be eccentric, which will affect the signal in the LISA frequency band when dynamically formed binaries should not yet been circularized. 

We will show in particular the accumulated number of BBHs per frequency bin can be a very useful probe of eccentricity distribution and thus of formation channels, even in the absence of eccentric templates.  We also provide a simple formula for the signal-to-noise ratio (SNR) of eccentric binaries that holds well for all eccentricity between 0 and 1 without explicit summation over higher harmonic components.

The most important influence on the number count is that the peak frequency with which GWs are emitted depends on eccentricity. In contrast to past predictions for multichannel observations \cite{Sesana:2016ljz,Gerosa:2019dbe}, any binaries observed with nonzero eccentricity at LIGO would not radiate at a measurable level in the LISA window at all  due to their high eccentricity at formation. Without eccentric templates, the distribution would also not provide LISA binaries as the precursors of the LIGO distributions from most dynamical channels (even those for which LIGO eccentricity is too small to be measured) and some of those dynamically formed binaries will emit chiefly at frequencies between those of the LIGO and LISA bands. 

Furthermore, we will demonstrate that the signal-to-noise ratio (SNR) of BBHs (even those measured with eccentric templates) decreases with eccentricity, further reducing the number of accessible dynamically formed binaries.  

This reduction is sometimes partially compensated by the fact that for fixed local merger rate, the number density of inspiraling BBHs per frequency range is enhanced by eccentricity. (See also \cite{Fang:2019dnh}). 

The net effect of these factors is that eccentricities affect the observable number of BBHs so that the absence of detected binaries with circular templates will have much greater discriminatory power than would be naively anticipated.  It is not merely that without eccentric templates we don’t see eccentric binaries. We are making a much stronger statement that for sufficiently large eccentricity in the LIGO window, entire populations will be lost, even in some cases with eccentric templates.

\emph{Evolution of eccentric binaries.} We briefly review the evolution of eccentric binaries and their GWs. Eccentric binaries are in general formed in dynamical channels in which the tidal influence of ambient masses is important. However, when these binaries enter the LISA band, their separation has been reduced enough that these gravitational disturbances to the binary orbit should be negligible, with the exception of possible but infrequent nonperturbative influences. Because most LISA binaries are already outside the range of significant tidal influences,  we treat them as isolated during the time in the LISA window and use the leading PN approximation with quadrupole radiation only, which holds well during the early inspiral of BBHs potentially visible to LISA.  

\begin{figure} 
\centering
\includegraphics[width=0.44\textwidth]{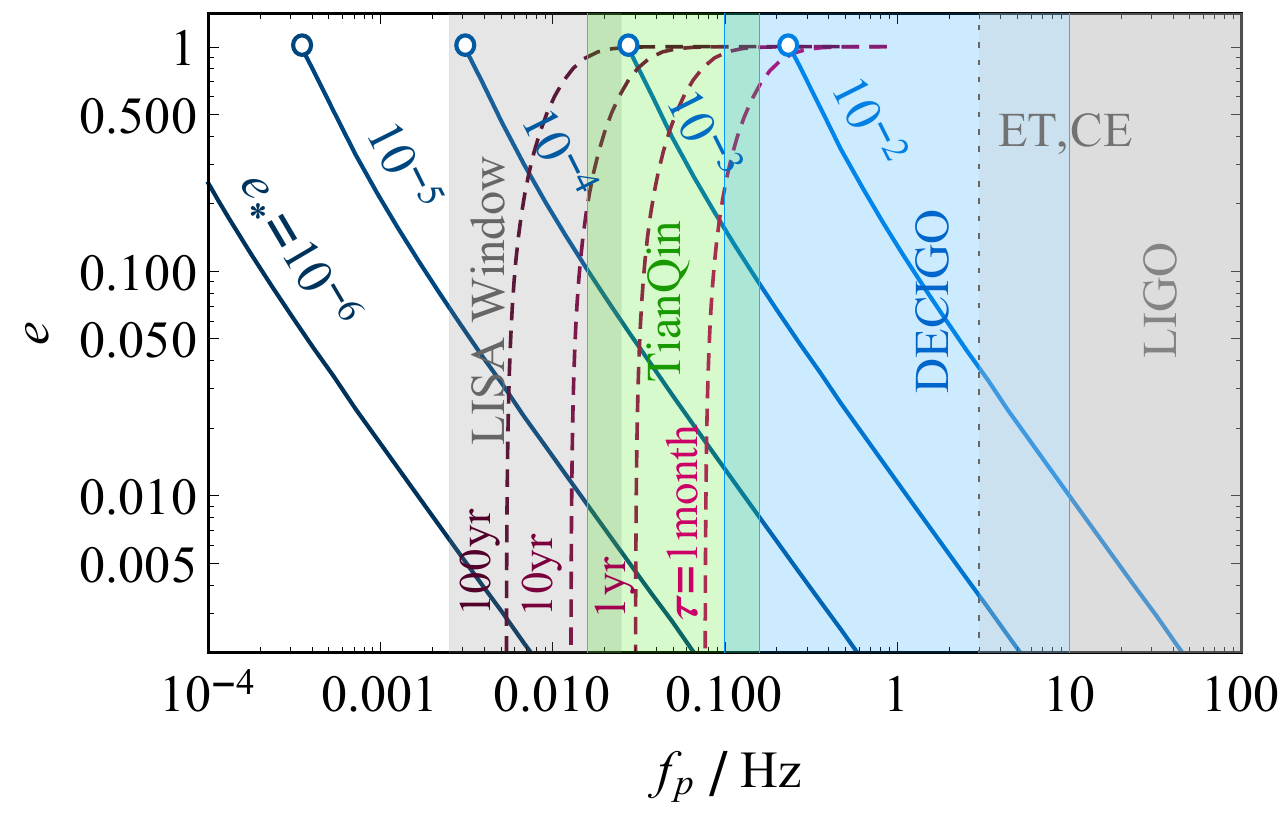}  
\caption{The binary eccentricity $e$ as a function of the peak GW frequency $f_p$. The five blue solid curves correspond to five reference values $e_*=10^{-n}$ $(n=2,3,4,5,6)$ at $f_{p*}=10$Hz, respectively. The four dashed magenta curves show the time $\tau$ to coalescence of binaries with $m_1=m_2=30M_\odot$. The shaded strips show the frequency ranges covered by several GW telescopes. } 
\label{fig_he}
\end{figure}

\begin{figure} 
\centering
\includegraphics[width=0.4\textwidth]{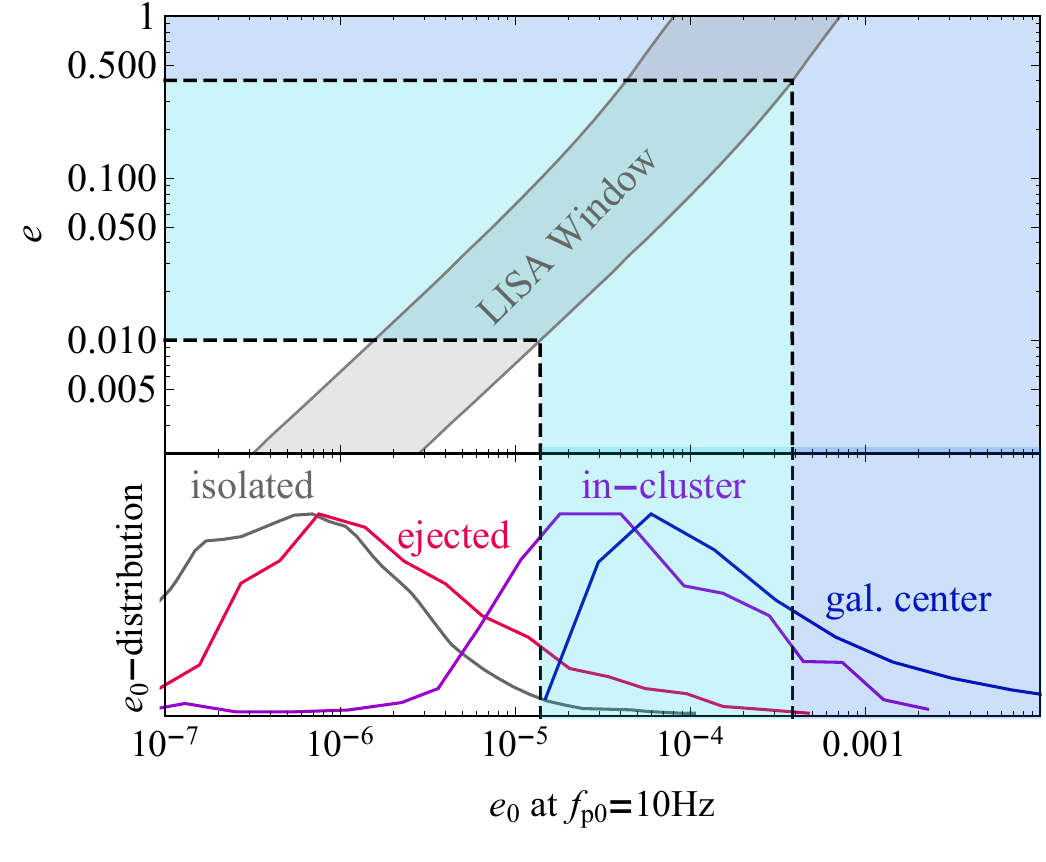}  
\caption{UPPER: The eccentricity $e$ in the LISA window (grey strip, same as in Fig.\ \ref{fig_he}) versus the eccentricity $e_*$ at $f_{p*}=$10Hz. BBHs to the lower-left of the black dashed lines could be seen in LISA if LISA is able to measure $e$ up to 0.01 or 0.4, respectively. LOWER: Eccentricity distributions from several channels at 10Hz. The four curves corresponds to the isolated channel \cite{Nishizawa:2016eza}, the ejected binaries from globular clusters and the in-cluster mergers \cite{Rodriguez:2018pss}, and binaries from galactic centers \cite{Randall:2018nud}. All curves are normalized at their peak values and the overall heights do not represent relative fractions of channels. } 
\label{fig_ee}
\end{figure}

We consider a binary made of masses $m_1$ and $m_2$, with total mass $m=m_1+m_2$ and reduced mass $\mu=m_1m_2/m$. We will also use the chirp mass $m_c\equiv \mu^{3/5}m^{2/5}$. The orbit of the binary is elliptical in general, with semi-major axis $a$ and eccentricity $e$.

The back reaction of GWs reduces and circularizes the binary orbit. At the quadrupole level, the evolution of $a(t)$ and $e(t)$ are described by the Peters' equations \cite{Peters:1964zz},
\bgs
\label{peters}
\begin{align}
\frac{\di a}{\di t} =&-\frac{64}{5}\FR{G^3\mu m^2}{c^5a^3}\FR{1+\fr{73}{24}e^2+\fr{37}{96}e^4}{(1-e^2)^{7/2}} ,\\
\frac{\di e}{\di t} =&-\frac{304}{15}\FR{G^3\mu m^2}{c^5a^4}\FR{e(1+\frac{121}{304}e^2)}{(1-e^2)^{5/2}},
\end{align}
\eds
where $G$ is  Newton's constant and $c$ is the speed of light. Eliminating $t$ from Peters' equation, we get a relation between $a(t)$ and $e(t)$ for a binary with initial value $(a_0,e_0)$,
\bge
\label{ge}
  \FR{a}{a_0}=\FR{\mathcal{G}(e)}{\mathcal{G}(e_0)}, ~~\mathcal{G}(e)\equiv\FR{e^{12/19}}{1-e^2}\bigg(1+\FR{121}{304}e^2\bigg)^{870/2299}.
\ede
The GW radiation from elliptical systems will have many harmonic components at integer multiples of the orbital frequency $f_\text{orb}=(2\pi)^{-1}\sqrt{Gm/a^3}$. Circular binaries have only the second harmonic $f_\text{gw}=2f_\text{orb}$ due to the quadrupole nature of the GW radiation. Elliptical binaries' GW spectrum will peak at $f_p$ which we call the peak frequency and is given by \cite{Randall:2017jop}
\bge
  \label{fp}
  f_p\simeq\FR{\sqrt{Gm}(1+e)^\ga}{\pi [a(1-e^2)]^{3/2}},~~~~~\ga= 1.1954.
\ede
Importantly, nonzero eccentricity shifts the peak of the GW spectrum towards higher frequency. For large $e\lesssim 1$, we have $f_p\simeq \pi^{-1}\sqrt{Gm/a_p^3}$ where $a_p=a(1-e)$ is the periapsis distance. Consequently, $f_p\gg f_\text{orb}$ when $e$ is large, so it is possible that $f_p$ is in the LISA frequency range while $f_\text{orb}$ is inaccessible to LISA.  We see that when studying the evolution of eccentric binaries relating to LISA observations, it is generally more useful to use $f_p$ and $e$ as independent variables instead of $a$ and $e$. 

Similar to (\ref{ge}), we can find a relation between the evolution of $e$ and $f_p$ by eliminating $a$ from (\ref{ge}) and (\ref{fp}). For a binary with eccentricity $e_*$ and peak frequency $f_{p*}$ at some moment, we have,
\begin{align}
\label{he}
  &\FR{f_p}{f_{p*}}=\FR{\mathcal{H}(e)}{\mathcal{H}(e_*)}, &&\mathcal{H}(e)\equiv\FR{(1+e)^\ga}{[(1-e^2)\mathcal{G}(e)]^{3/2}}.
\end{align}
The function $\mathcal{H}(e)$ has the important property that it reaches a finite constant $\mathcal{H}(1)\simeq 1.89$ when $e\to 1$. This means that for a binary with nonzero eccentricity $e_*$ at a fixed frequency $f_{p*}$, the frequency $f_p$ at earlier times would never be smaller than a cutoff frequency $f_\text{min}\equiv [\mathcal{H}(1)/\mathcal{H}(e_*)]f_{p*}$. The cutoff frequency $f_\text{min}$ depends only on $e_*$ and $f_{p*}$ but not on the masses of the binary. Note this result differs from Ref. \cite{Nishizawa:2016eza}, which assumed the GW frequency $f_\text{GW}=2f_\text{orb}$ instead of $f_p$, which is approximately correct for the low eccentricity binaries they considered but would be inadequate for larger $e$ values.

Fig.\ \ref{fig_he} shows (\ref{he}), choosing $f_{p*}=10$Hz at the lower end of the LIGO band for different $e_*$ at 10Hz.  We see that, for instance, a binary with $e_*=10^{-3}$ in LIGO (a value not measurable in LIGO but a value motivated by dynamical predictions) would never radiate measurable GWs below 0.02Hz. So for example if all binaries had  eccentricity higher than $10^{-3}$ at 10Hz, LISA would see no binaries below 0.02Hz.  

Although extreme, this shows how it could in principle be possible to infer the eccentricity distribution by counting the binary number measured in LISA, even without measuring eccentricity directly in either detector (in the case of LISA not seeing events at all). In Fig.\;\ref{fig_ee} we show the coverage of LISA in $e_*$ in light of the predicted $e_*$ distribution from several formation channels.

%If LISA only uses circular templates and is blind to all highly eccentric binaries, then no binaries would be seen below $\sim0.1$Hz, almost the upper end of the LISA band.

\emph{Number density of BBHs.}  Contrary to the transient nature of BBH mergers in LIGO, most stellar mass BBHs in LISA emit GWs with slowly varying frequency. Because of the eccentricity-dependent evolution in frequency between the LISA and LIGO bands conveyed in Fig.~\ref{fig_he}, the number density of stellar mass BBHs per frequency range $\di n/\di f_p$  conveys information about the formation channel and corresponding eccentricity distribution. Here $n$ is the comoving number density of the BBHs, and we are using the peak frequency $f_p$ as the variable to account for elliptical orbits.

The number density $\di n/\di f_p$ can be inferred from the local merger rate $\mathcal{R}=\di n/\di t$ by the chain rule, $\di n/\di f_p=\mathcal{R}\dot f_p^{-1}$. The rate of ``chirping'' $\dot f_p=(\pd f_p/\pd a)\dot a+(\pd f_p/\pd e)\dot e$. With (\ref{peters}) and (\ref{fp}), we get,
\begin{align}
\label{dtdfp}
  \FR{\di t}{\di f_p}=&~\FR{5c^5}{96\pi^{8/3}}(Gm_c)^{-5/3}f_p^{-11/3}\mathcal{F}(e),
\end{align}
\begin{align}
  \label{fe}
  \mathcal{F}(e)\equiv&~\FR{(1+e)^{8\ga/3-1/2}}{(1-e)^{3/2}}\big[(1+e)(1+\fr{7}{8}e^2)\n\\
 &~-\fr{\ga}{288} e(304+121e^2)\big]^{-1}.
\end{align}
The function $\mathcal{F}(e)$ describes the correction from $e\neq 0$. One should be careful when using (\ref{dtdfp}) because the eccentricity $e$ in (\ref{dtdfp}) is not a constant and changes with time, or equivalently, with $f_p$. The relation between $e$ and $f_p$ is given in (\ref{he}). So we should replace $e$ in (\ref{fe}) with $e=\mathcal{H}^{(-1)}[(f_p/f_{p*})\mathcal{H}(e_*)]$ where $e_*$ is the eccentricity at a fiducial frequency $f_{p*}$ which we choose to be $10$Hz.

Note that $\mathcal{F}(e)>1$ when $e>0$ and so finite eccentricity enhances BBH number density per frequency range (for frequencies with nonzero contribution) compared to circular ones. Qualitatively one can understand this enhancement by looking at the integrated number density $n=\mathcal{R}\int\di f_p(\di t/\di f_p)\sim \mathcal{R}\tau$ where $\tau$ is the lifetime of BBHs. The lifetime of a binary starting from initial $(a_0,e_0)$ scales like $\tau\sim a_0^4\ep_0^{7/2}\sim f_p^{-8/3}\ep_0^{-1/2}$ \cite{Randall:2018nud}. That is, the lifetime increases with eccentricity for fixed $f_p$. Consequently, when eccentricity is increased, more binaries are needed to achieve the same merger rate.

It might seem counterintuitive for eccentricity to decrease the merger rate. We emphasize that this is true only for fixed initial peak frequency $f_p$. If one fixes the initial $2f_\text{orb}$ instead of $f_p$, then the merger time for eccentric orbits is lower. 

Combined with the fact that BBHs circularize during inspiral, the above discussion shows that if these were the only effects, the relative fraction of eccentric binaries would increase at lower peak frequency for frequencies above the frequency cutoff -- the frequency $f_\text{min}$ below which the eccentric binary could never reach. However,  SNR decreases with eccentricity and furthermore the peak frequency for large $e$ will move out of the LISA range — both of which can lead to very different frequency distributions for eccentric orbits as we now explore.

The distribution of BBHs can be written as a function of the chirp mass $m_c$, the distance from the observer $r$, the peak frequency $f_p$, and the eccentricity $e_*$ at a fiducial frequency $f_{p*}$ which we choose to be 10Hz, 
\bge
\label{dN}
  \FR{\di^4N}{\di m_c\di r\di f_p\di e_*}=4\pi r^2\mathcal{R}\FR{\di t}{\di f_p}f(m_c,e_*).
\ede
We have neglected cosmic expansion since most LISA BBHs will have small redshift. 

The distribution $f(m_c,e_*)$ depends on the formation channel. Isolated binaries from common envelope channels should be almost circular \cite{Nishizawa:2016eza} so that most such binaries have $e_*<10^{-6}$. On the other hand, dynamical channels generally lead to larger $e_*$ roughly ranging from $10^{-6}$ to $10^{-3}$, as shown in Fig.\ \ref{fig_ee}. Remarkably, we see that LISA has just the right frequency band to have the potential to distinguish predictions with $10^{-6}\lesssim e_* \lesssim 10^{-3}$. Below we shall show that LISA does not even have to measure $e$ very accurately; a counting of BBH number in every frequency bin could reveal enough information to distinguish dynamical from isolated channels. 

To describe how this is achieved, we need to include the effect of nonzero eccentricity on a BBH's signal-to-noise ratio (SNR) in LISA.

\begin{figure} 
\centering
\includegraphics[width=0.23\textwidth]{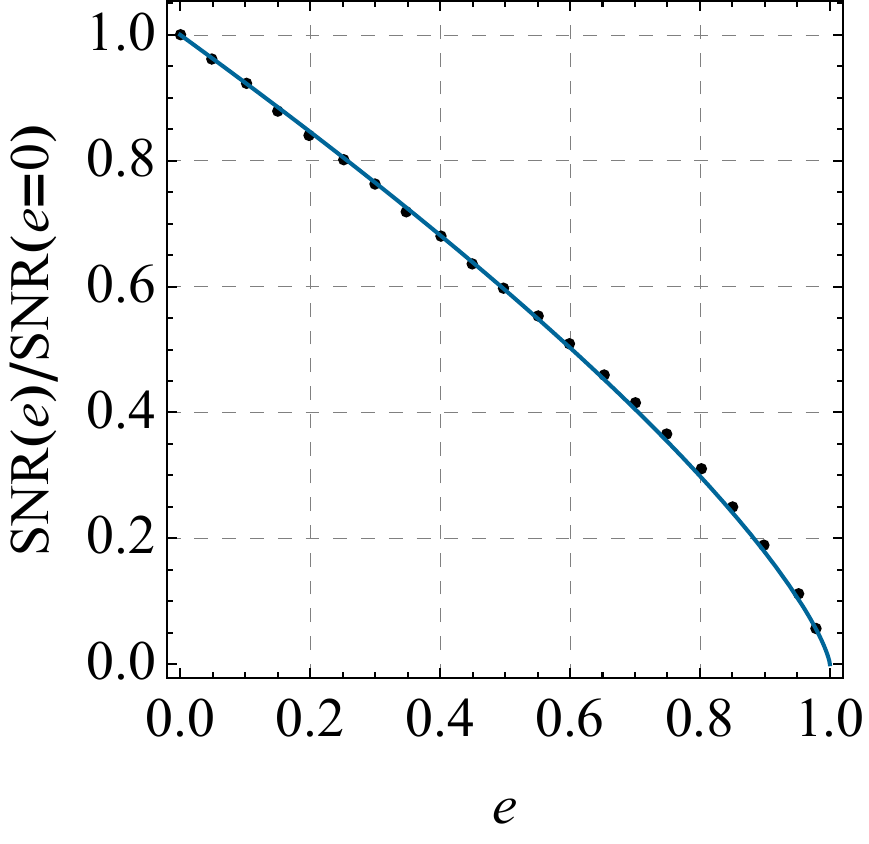}  
\includegraphics[width=0.23\textwidth]{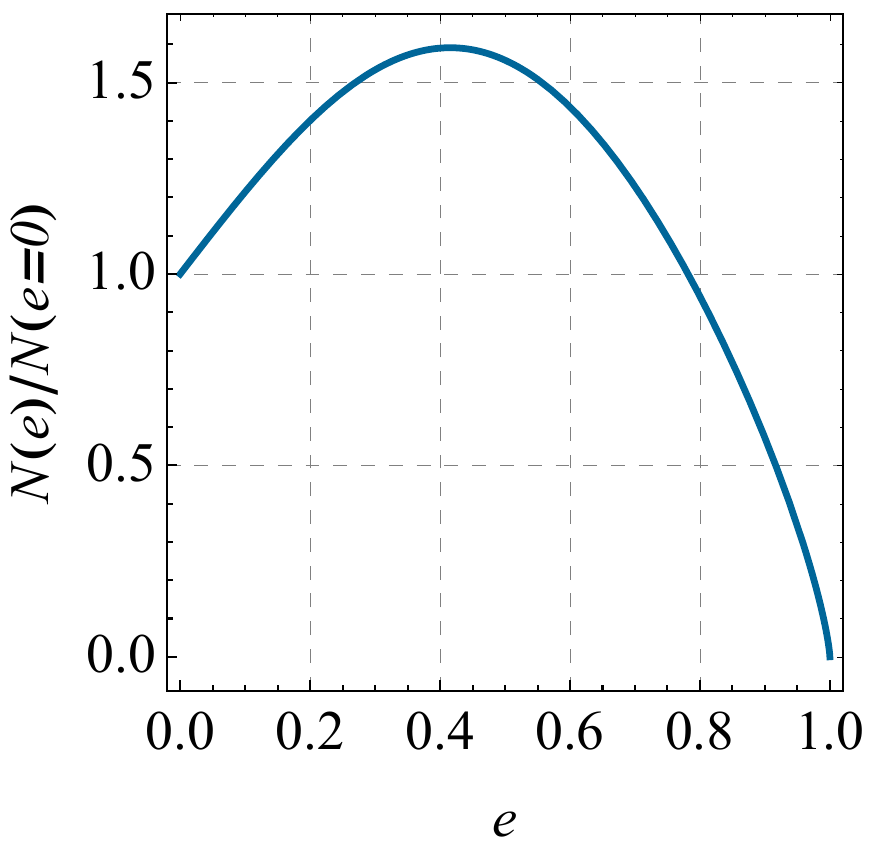}  
\caption{LEFT: The SNR of a non-chirping eccentric binary as a function of eccentricity $e$, with peak frequency $f_p$ and all other parameters fixed. The black dots are calculated from summation over harmonics and the blue curve shows the simplified formula (\ref{snrnonchirp}). RIGHT: The relative enhancement/suppression of expected number of BBHs in LISA due to finite eccentricity.} 
\label{fig_snr}
\end{figure}

\begin{figure*}[t]
\centering
\includegraphics[width=0.95\textwidth]{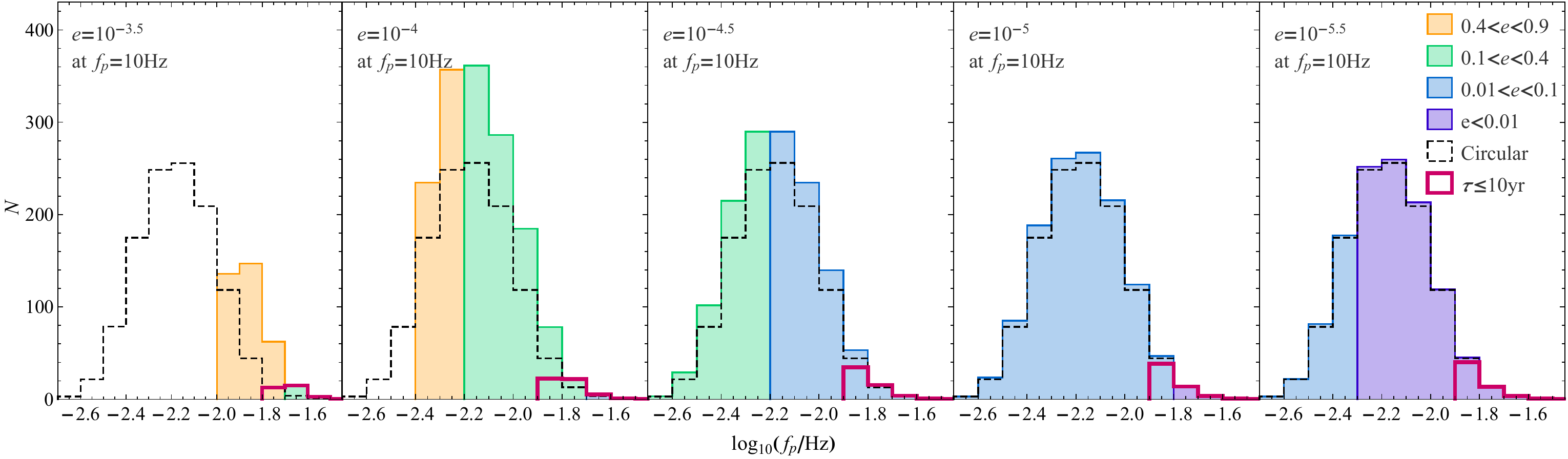}  
\caption{The number of resolvable ($\varrho>8$) BBHs in LISA with N2A5 configuration \cite{Klein:2015hvg} and 10yr observation. In all panels, we use dashed black lines to show a circular distribution with $e_*=0$, which serves as a basis to which we compare number distribution with finite $e_*$. In each panel, we choose a different $e_*$ at 10Hz ranging from $10^{-3.5}$Hz to $10^{-5.5}$Hz. The purple, blue, green, and orange shadings correspond to $e_\text{cut}=0.01,0.1,0.4,0.9$, respectively. The binaries enclosed by magenta lines merge in 10 years so are possible for joint detection with ground GW telescopes.} 
\label{fig_nf}
\end{figure*}

\emph{LISA SNR of eccentric BBHs.} The SNR $\varrho$ of eccentric BBHs can be calculated as
\bge
\label{snr}
  \varrho^2=4\sum_{n}\int\FR{h_n^2(f_n(t))}{S_N(f_n(t))}\di t,
\ede
where the summation is over all harmonic component of GWs. For highly eccentric BBHs one has to sum over a large number of harmonics which can be numerically challenging. However, the noise strain $S_N(f)$ does not vary a lot within the width of the GW spectrum, so a good approximation is to pull $S_N$ out of the summation, so we get $\sum h_n^2=\la h^2\ra\equiv h_c^2$, which is simply the amplitude of GW radiation averaged over one orbit. To see how this quantity depends on  eccentricity, first consider the large $e$ limit where we keep track of $\ep=1-e^2$ factors. Then the GW amplitude is proportional $\ddot M$ with $M\propto (a \ep)^2$, the mass quadrupole of the binary. To take the time derivative, we use the fact that $\di/\di t=\dot\psi(\di/\di\psi)$ where $\psi$ is the true anomaly of the binary orbit on which the mass quadrupole has sinusoidal dependence, and $\dot\psi\propto (a\ep)^{-3/2}$. Therefore,
\bge
  \la h^2\ra \propto \la \ddot M^2\ra\propto \omega_0\int_0^{2\pi}\di\psi\,\dot\psi^{-1}\ddot M^2\propto a^{-2}\ep^{-1/2}.
\ede
Now using $f_p\propto (a\ep)^{-3/2}$, we have $\la h^2\ra\propto f_p^{4/3}\ep^{3/2}$. There we see  if we treat the SNR as a function of $f_p$ and $e$, then it scales with $\ep$ like $\ep^{3/4}$.

In fact a simple formula turns out to well approximate the sum over harmonics for any value of  eccentricity $0\leq e<1$:  
\bge
  h_c^2(f_p,e)=h_c^2(f_p,e=0)\cdot(1-e)^{3/2}.
\ede 
Therefore, for binaries with little chirping during the whole observation time so that $f_p$ and $e$ are relatively constant, we have,
\bge
\label{snrnonchirp}
  \varrho(f_p,e)=\varrho(f_p,e=0)\cdot(1-e)^{3/4}.
\ede 
For chirping binaries, we can extend to the following generalized expression,
\begin{align}
  \label{snrsimp}
  \varrho^2(f_p,e)=4\int\di t\,\FR{h_c^2(f_p(t),e=0)}{S_N(f_p(t))}\big[1-e(t)\big]^{3/2}.
\end{align}
Here $f_p(t)$ and $e(t)$ should be calculated using  (\ref{peters}) and (\ref{fp}). In this way we can avoid the summation over GW harmonics. We compare the results from the simplified equation (\ref{snrnonchirp}) and from the original one (\ref{snr}) in the left panel of Fig.\;\ref{fig_snr} and find very good agreement for all eccentricities.  

(\ref{snrnonchirp}) shows that a nonzero eccentricity will decrease the SNR for a binary with fixed peak frequency. One can again understand this suppression of SNR by noting that the eccentricity lengthens the lifetime $\tau$ for fixed $f_p$ and thus suppresses the radiation power $\dot E\sim \tau^{-1}$.

\begin{figure}[t]
\centering
\includegraphics[width=0.4\textwidth]{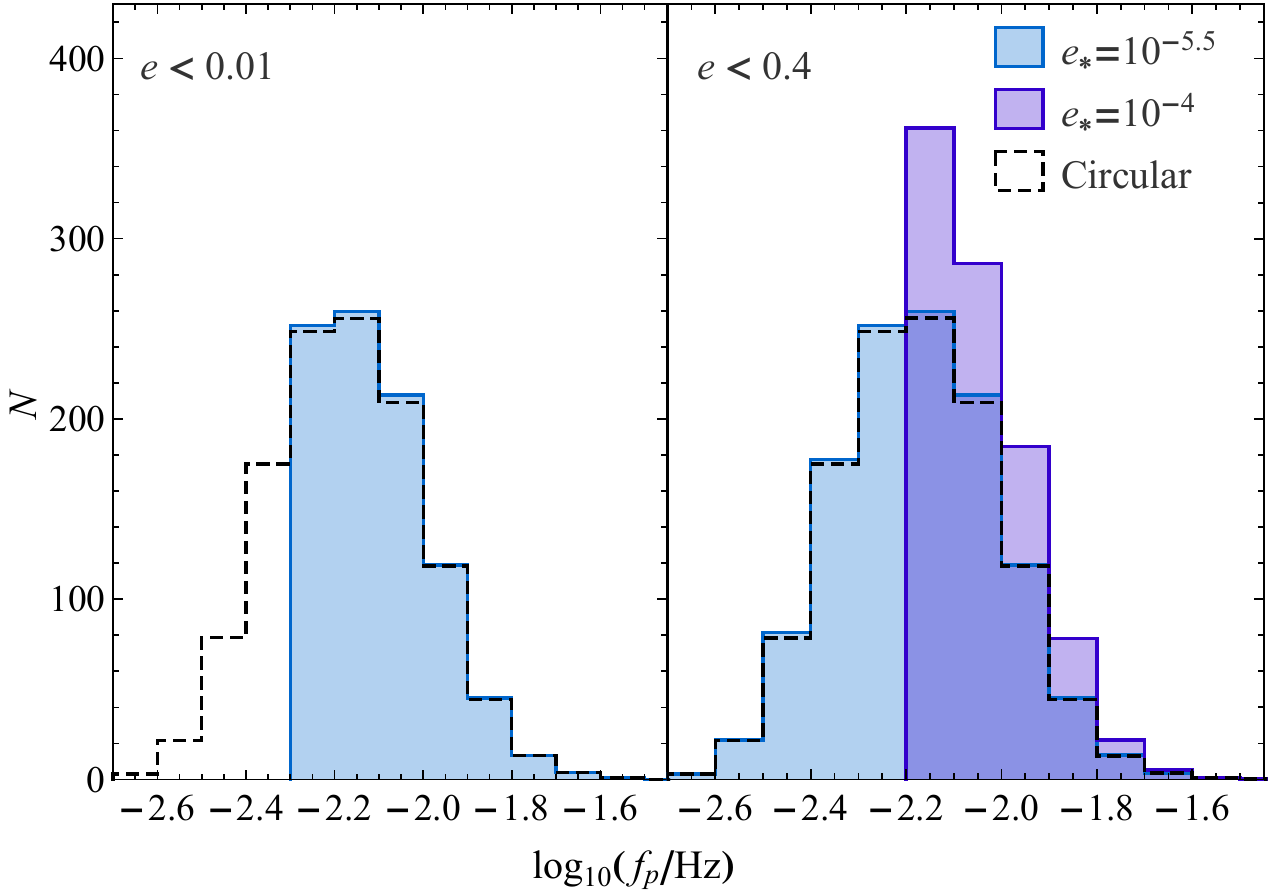}  
\caption{Rearrangement of histograms in Fig.\;\ref{fig_nf} to highlight the effects of eccentricity $e_*$ on the number of resolvable binaries. The left (right) panel shows the resolvable number with eccentricity in LISA smaller than 0.01 (0.4). The blue and purple shadings correspond to $e_*=10^{-5.5}$ and $10^{-4}$, respectively. } 
\label{fig_nf2}
\end{figure}

\emph{Number distribution in the frequency domain.} With the  number density (\ref{dN}) and the SNR (\ref{snrsimp}) of eccentric BBHs, we can now evaluate the number of resolvable BBHs in LISA. Since SNR scales like $(1-e)^{3/4}/r$ with $r$ the source distance, finite $e$ will decrease the total volume reachable by LISA as $(1-e)^{9/4}$. On the other hand, the eccentricity increases number density by a factor of $\mathcal{F}(e)$ defined in (\ref{fe}). Therefore, we see that the total number of resolvable BBHs scales with $e$ according to $(1-e)^{9/4}\mathcal{F}(e)$. We plot this combination in the right panel of Fig.\ \ref{fig_snr}. We see that the total number is enhanced for $0<e\lesssim0.8$ but suppressed for high $e\gtrsim 0.8$. We note that the enhancement of resolvable BBHs is most prominent for $e\sim 0.4$, which happens to be around the largest $e$ reached by state-of-the-art eccentric templates.  Through the relation (\ref{he}) between $e$ and $f_p$, the enhancement around $e\sim 0.4$ is translated to the enhancement of resolvable BBHs in the LISA band ($f_p\sim0.01$Hz) for dynamically formed BBHs with $e_*\sim10^{-4}$ at $f_p=10$Hz. On the other hand, the suppression for $e>0.8$ is translated to the reduction of binaries numbers at low $f_p$ compared with a circular distribution. This leads to the most important effect, which is that signals associated with finite eccentricity will be concentrated at higher frequencies, potentially outside the LISA band.

To calculate the number of resolvable BBs, we assume 10yrs of LISA observations, and use $\varrho\geq\varrho_\text{min}=8$ as a criterion of resolvability. To illustrate the effect of eccentricity, we take $m_1=m_2$, and also the merger rate $\mathcal{R}\simeq 50$Gpc$^{-3}$yr$^{-1}$ inferred from LIGO/Virgo observations \cite{LIGOScientific:2018jsj}, which gives $\mathcal{R}=53.2^{+58.5}_{-28.8}\text{Gpc}^{-3}\text{yr}^{-1}$ assuming a mass function $p(m_1)\propto m_1^{-2.3}$ within $(5M_\odot,50M_\odot)$.
 
We then calculate the total number of resolvable BBHs in the frequency bin $[f,f+\Delta f]$. 
\begin{align}
  &N(f,\Delta f,e_\text{cut})=\mathcal{R}\int_f^{f+\Delta f}\di f_p\int\di r\di m\, p(m) \,   4\pi r ^2\FR{\di t}{\di f_p}\n\\
  &~\times\theta\big[\varrho(r,f_p,m_c,e_*)-\varrho_\text{min}\big]\theta\big[e_\text{cut}-e(f)\big],
\end{align}
where we introduced two $\theta$-function cutoffs. The first is to select BBHs with SNR larger than $\varrho_\text{min}=8$, and the second characterizes our limited ability to see very eccentric BBHs, due to template limitations. One can get information about eccentricity by measuring $N(f,\Delta f,e_\text{cut})$ even with small $e_\text{cut}$.  We illustrate this in Fig.\ \ref{fig_nf} with several examples of $N(f,\Delta f,e_\text{cut})$ as function of $f$. We choose log-uniform bin width $\Delta f=10^{0.1}f$.  

We see that without eccentric templates we would miss entire populations of binaries unless $e_*$ is sufficiently small. Of course it is clear that if the template doesn’t cover the eccentricity we won’t see it. The new wrinkle here is the relationship between the black holes predicted for LIGO and where in frequency/eccentricity space they are expected in LISA. Many of the expected binaries would be lost in the LISA window without eccentric templates. 

%??we need to be careful throughout to say events in lisa band—they are notis our point

 For example if LISA is blind to binaries with $e>0.01$ it will see no binaries at all for distributions with $e_*\geq 10^{-5}$. In the range of $10^{-5}\lesssim e\lesssim 10^{-6}$, LISA will find a significant reduction in $N(f,\Delta f,e_\text{cut})$ at lower frequencies. The message is that if LISA sees significantly less BBHs than naively expected, one should search for BBHs with eccentric templates. 

We demonstrate this in Fig.\ \ref{fig_nf}.  The  blue, green, and orange shadings show how many more BBHs one can see if the largest observable eccentricity is raised to 0.1, 0.4, and 0.9, respectively. By including these modest $e$ values, one can probe $e_*$ up to $\order{10^{-4}\sim 10^{-3}}$. We mentioned earlier that  dynamical channels will mostly lead to  distributions of $e_*$ peaked between $10^{-6}$ and $10^{-3}$, while isolated channels mostly produces binaries with $e_*<10^{-6}$. This result tells us that LISA can in principle distinguish these classes of channels by doing an $N(f,\Delta f,e_\text{cut})$ measurement. A circular-only measurement should definitively distinguish $e_*$ being above or below $\order{10^{-5}}$ and thereby tell the difference between dynamical channels and isolated channels. By further applying eccentric templates capable of measuring $e$ up to 0.1 or 0.9, one can further distinguish among different sub-classes of dynamical channels with different peak $e_*$ (and of course verify this interpretation).

In Fig.\ \ref{fig_nf2}, we further contrast $N(f,\Delta f,e_\text{cut})$ from two channels with $e_*=10^{-5.5}$ and $10^{-4}$, respectively. The left panel shows what LISA will see using circular templates only. In this case LISA will see no binaries from the $e_*=10^{-4}$ channel and a reduced number of binaries at lower frequencies from the $e_*=10^{-5.5}$ channel. The right panel shows the resolvable number if templates up to $e=0.4$ are used.

In each panel of Fig.\ \ref{fig_nf}, we also show the binaries merging within ten years in the LIGO band that could conceivably be targets of multiband GW observation. All these binaries will have tiny eccentricity in the LIGO band. 
We can hope for multiband observations with either eccentric templates or intermediate frequency observations.  Very high eccentricity templates remain challenging. 
With templates measuring $e$ to 0.4 however, we can observe the majority of stellar-mass black holes formed in either isolated or dynamical channels. In particular, for the distribution with $e_*=10^{-4}$ predicted by several dynamical channels (cf.~Fig~\ref{fig_ee}), templates up to $e\sim0.4$ will allow us to see more BBHs in the LISA band than expected form circular distribution. This is most clearly seen from the second panel of Fig.~\ref{fig_nf}. 

%??again commewtn on when F gives us enhancement

%Without such templates, the only way to observe the dynamically formed black hole pairs is with higher frequency detectors, as illustrated in Fig.~\ref{fig_he}. If we see a deficit in binary observations at LISA, multiband observations in the intermediate frequency regions from 0.1 to 10Hz become much more interesting. Multiband observations with circular templates  require sub-Hertz GW telescopes such as DECIGO and Tianqin, and the next generation ground-based interferometer can also help by going to lower frequency, such as planned for the Einstein Telescope (ET) and the Cosmic Explorer (CE).  

We emphasize that varying the mass function $p(m)$  affects only the overall number of resolvable BBHs, but not the distribution $\di N/\di f$. In fact, the number of resolvable BBHs is very sensitive to the BBH mass function which we don't know well enough. This uncertainty will decrease with more observations in the future.  Nonetheless, apart from the overall scaling,  the number in each bin $N(f,f+\Delta f)$ as a function of $f$ is not sensitive to the mass function. However, the $e_*$-distribution from various dynamical channels could correlate with the mass $m$. We leave a more systematic study with different $e_*$-distributions to the future.

It is of interest to compare our results with previous studies \cite{Nishizawa:2016eza}. While we reach a similar conclusion that LISA will see  fewer binaries for dynamical channels, the underlying reasons are very different. \cite{Nishizawa:2016eza} claimed that the eccentricity will boost the merger rate and thus reduced the number density of inspiraling binaries for fixed merger rate, which in turn led to a reduced predicted number of observable binaries. This claim implicitly assumed GW frequency being twice the orbital frequency, $f_\text{GW}=2f_\text{orb}$. We emphasized in this paper that we should use the peak frequency $f_p$ instead of $2f_\text{orb}$. Based on this, we showed that the eccentric binaries have a cutoff frequency $f_\text{min}$ that depends only on its eccentricity $e_*$ at a reference frequency $f_{p*}$. Above $f_\text{min}$, we show that eccentricity actually \emph{increases} the number density. However, we also showed that eccentricity reduces SNR even with appropriate eccentric templates. Combining all these effects and accounting for the eccentricity-dependence of the peak frequency, we showed that the number of observable binaries will be decreased for dynamical channels if we use circular templates only. But, with appropriate eccentric templates, the number of observable binaries in LISA in the case of dynamical formation can be either larger or smaller than the isolated channel, which can be seen from Fig.\ \ref{fig_nf}.

\emph{Multiband observation of eccentric LIGO binaries.} We have shown that binaries with eccentricity larger than $10^{-3}$ in LIGO at 10Hz will not be observable in LISA. A multiband observation for such binaries would require GW telescopes working at sub-Hertz band, such as DECIGO \cite{Kawamura:2011zz} and TianQin \cite{Luo:2015ght}. The next generation ground-based interferometer can also help by going to lower frequency, such as planned for the Einstein Telescope (ET) \cite{Punturo:2010zz} and the Cosmic Explorer (CE) \cite{Evans:2016mbw}. We illustrate their coverages in the frequency domain in Fig.\ \ref{fig_he}. Interestingly, binaries with $e_*$ lying between $10^{-3}$ and $10^{-2}$ will be essentially circular in LIGO while invisible in LISA. Therefore, sub-Hertz telescope would be the only way to see their eccentricity. In addition, binaries with observably large eccentricity in LIGO ($e_*>0.01$) are typically from non-perturbative dynamical process. Observing such binaries at sub-Hertz band would provide the only way to see the earlier evolution of such binaries before the merger, and would provide insights into the nature of the non-perturbative dynamical process.

Given that multiband observations for black holes being eccentric in LIGO will be unlikely,  it is interesting to extend the study to heavier binaries which could in principle be observed in both LISA and LIGO. However the total number of observable binaries will be limited by non-observation at LIGO so far. Taking $m_1=m_2=100M_\odot$ as an example, no observation in LIGO O1 data puts an upper limit of merger rate for this mass as $\mathcal{R}<2$Gpc$^{-3}$yr$^{-1}$ \cite{Abbott:2017iws}, which would be further lowered if also including O2 data. Then the total number of such heavy binaries in LISA with $\varrho>8$ in 10yr would be at most a few.

\emph{Discussion.} We have shown that LISA has just the right frequency range to distinguish among different formation channels with a number count in frequency bins. Although eccentric templates will help distinguish precise formation channels, because of the absence of low frequency eccentric mergers,  one can even probe eccentricity distributions by measuring circular binaries alone. We leave to future work  full statistical analysis of BBH distribution of mixed origins.

We also showed that binaries with $e_*>10^{-3}$ at 10Hz are not observable in LISA. This of course  includes all LIGO binaries with observably large eccentricity ($e_*\gtrsim 0.01$). Multiband observations of such binaries could however be possible with GW telescopes working between the LISA and LIGO bands.
Although the  expected number of observable binaries depends heavily on the mass function, we emphasize that this dependence will mostly affect the overall number but not the distribution in frequency.

We see from Fig.\ \ref{fig_ee} that there is still quite large degeneracy among formation channels even we can measure the eccentricity distribution perfectly. If future measurements show that a significant fraction of binaries are formed dynamically,  it will be desirable to find ways to further differentiate various dynamical channels, such as to study eccentricity in correlation with other parameters (e.g.\ binary mass \cite{Randall:2018nud}) or other type of signatures, such as binaries' barycenter motion \cite{Randall:2018lnh,Robson:2018svj,Wong:2019hsq} or tidal-induced eccentricity oscillations \cite{Randall:2019sab,Hoang:2019kye}.
   
\begin{acknowledgements} 

We thank Curt Cutler and the participants on the LISA White paper on multiband observations for stimulating this investigation.
LR was supported by an NSF grant PHY-1620806, the Chan Foundation, a Kavli Foundation grant ``Kavli Dream Team,'' the Simons Fellows Program, 
the Guggenheim Foundation, and a Moore Foundation Fellowship.
\end{acknowledgements}


%\end{fmffile}
\end{document}